\documentclass[final,5p,times,preprint]{elsarticle}
\usepackage[utf8]{inputenc}
\usepackage{amsfonts}
\usepackage{amsmath}
\usepackage{amssymb}
\usepackage{bm}
\usepackage{lineno,hyperref}
\usepackage{comment}
\usepackage{xcolor}
\usepackage{booktabs}
\usepackage{multirow}
\modulolinenumbers[5]
\usepackage[normalem]{ulem}
\usepackage{url}

\newcommand{\vecr}{{\vec r}}

\newcommand{\psixprior}{\varphi^\mathrm{prior}_x} 
\newcommand{\psixpost}{\varphi^\mathrm{post}_x} 
 
\newcommand{\psixno}{\varphi^\mathrm{HM}_x} 

\newcommand{\normsig}{\frac{2\pi}{\hbar v_i} \rho(E_b)}

\journal{Physics Letters B}

\biboptions{comma,sort&compress}
\begin{document}

\begin{frontmatter}

\title{
Neutron-transfer induced breakup of the Borromean nucleus $^9$Be}


\author[FAMN]{G. Villanueva}
\author[FAMN]{A. M. Moro\corref{mail}}
\cortext[mail]{Corresponding author}
\ead{moro@us.es}

\author[FAMN]{J. Casal}
\author[Tongji]{Jin Lei}

\address[FAMN]{Departamento de F\'{\i}sica At\'omica, Molecular y Nuclear, Facultad de F\'{\i}sica, Universidad de Sevilla, Apartado 1065, E-41080 Sevilla, Spain}


\address[Tongji]{School of Physics Science and Engineering, Tongji University, Shanghai 200092, China.}

\begin{abstract}
We address the problem of evaluating neutron-transfer induced breakup cross sections caused by the Borromean nucleus \(^{9}\)Be, using the reaction \(^{197}\)Au(\(^{9}\)Be,\(^{8}\)Be)\({\rm^{198}Au}\) as a test case. This reaction was recently measured over a wide range of incident energies around the Coulomb barrier. To deal with the high density of $^{198}$Au states that can be potentially populated in this reaction, we employ the Ichimura, Austern, Vicent model, in which the spectrum of physical states for this system is replaced by the solutions of a complex n+$^{197}$Au potential, accounting effectively for the fragmentation of single-particle states into physical states. Furthermore, to account for the unbound nature of the emitted $^{8}$Be system, we employ a three-body model of $^{9}$Be. The calculated stripping cross sections are found to be in good agreement with existing data over a wide range of incident energies. The importance of taking into account the energy spread of the single-particle strength of the outgoing \(^{8}\)Be and the target-like residual nucleus is discussed. 

\end{abstract}

\begin{keyword}
inclusive breakup \sep neutron transfer \sep three-body \sep $^{9,8}$Be

\end{keyword}

\end{frontmatter}


\section{Introduction}\label{sec:intro}
Experiments with light weakly-bound stable nuclei (such as $^{6,7}$Li and $^{9}$Be) have shown a systematic suppression of  complete fusion (CF)  cross sections   (defined as capture of the complete charge of the projectile) of  $\sim$20-30\% compared to the case of tightly bound nuclei \cite{Das99,Tri02,Das02,Das04,Muk06,Rat09,LFC15} as well as to coupled-channels calculations including the coupling to low-lying excited states of the projectile and target \cite{Das99,Das02,Kum12,Zha14,Fan15}.  The effect has been attributed to the presence of strong competing channels, such as the breakup of the weakly bound projectile prior to reaching the fusion barrier, with the subsequent reduction of capture probability. This interpretation is supported by the presence of large $\alpha$ yields as well as target-like residues which are consistent with the capture of one of the fragment constituents  of the projectile, a process which is usually termed as {\it incomplete fusion} (ICF). 

In the case of $^{6,7}$Li reactions, a successful quantitative account of the CF cross section has been achieved using a model based on the Continuum-Discretized Coupled-Channels (CDCC) method \cite{Ran20,Cor20} as well as using a novel method in which the CF cross section is obtained from the reaction cross section, after subtracting the dominant direct channels \cite{Jin19a,Jin19b}. 

In the case of the $^{9}$Be projectile, measured above-barrier CF cross sections are found to represent $\sim$70\% of those expected for well-bound nuclei  \cite{Raf10}. Dedicated efforts have been devoted to identify the competing channels that are responsible for this apparent CF
suppression. A promising candidate is the breakup following neutron transfer, which has been found to represent a significant part of the reaction cross section, notably exceeding the direct breakup of the projectile. Other experiments with $^{9}$Be  have confirmed the dominant role of the one-neutron stripping cross section  \cite{gollan21,kaushik21}.





To investigate the impact of the neutron stripping channel on the CF cross sections, one should first be able to accurately predict the cross sections for the former.  These calculations face two main difficulties: (i) the outgoing $^{8}$Be core resulting from the one-neutron removal from the projectile is unbound, and (ii) the number of final states is very high, making it very difficult to resolve individual states. Moreover, even for those states that can be experimentally resolved, the spin-assignment and spectroscopic factors, which are key quantities required in standard distorted-wave Born approximation (DWBA), coupled-chanels Born approximation (CCBA) or coupled reaction channels (CRC) calculations, are poorly known for excitation energies above a few MeV. For example, the CRC calculations presented in \cite{gollan21} for the $^{197}$Au($^{9}$Be,$^{8}$Be)$^{198}$Au reaction limited the states of $^{198}$Au below 700~keV, leaving out a significant part of the bound-state spectrum of this nucleus ($S_n=6.5$~MeV).

In this work, we propose a new approach that aims to overcome the two aforementioned limitations. First, the unbound nature of the $^{8}$Be system is accounted for by using a three-body description of the $^{9}$Be projectile ($\alpha + \alpha +n$). This allowed the calculation of the required $\langle  {^8}\text{Be} | {^9}\text{Be} \rangle$ overlaps without resorting to an artificial n-$^{8}$Be potential. Second, the high density of states of the target nucleus is incorporated in the reaction using the Ichimura, Austern, Vincent (IAV) model for inclusive breakup reactions. In this model, the physical spectrum of the neutron+target system is replaced by an approximate description in terms of a complex neutron+target optical potential, whose imaginary part accounts for the fragmentation of the single-particle strength into the physical states. This avoids the need for the detailed knowledge of the spin-assignment and associated spectroscopic factors of the populated states and enables a tractable description of this otherwise extremely complex reaction process without resorting to arbitrary truncations of the spectrum of populated states. 

The paper is organized as follows. In Sec.~\ref{sec:calc} we give a short overview of the reaction formalism and of the three-body model adopted for the description of the $^{9}$Be projectile. In Sec.~\ref{sec:results} we present the results of the calculations for $^{197}$Au($^{9}$Be,$^{8}$Be)${\rm^{198}Au}$. Finally, the main conclusions of the work are given in Sec.~\ref{sec:sum}.

\section{Theoretical formalism}\label{sec:calc}
\subsection{Reaction framework}
We revisit the derivation of the IAV formula, with emphasis in the modifications required to account for the composite (and unbound) structure of the spectator system. A more detailed description of the method can be found in the original work of IAV \cite{IAV85} as well as in more recent applications~\cite{Jin15,Pot15,Car16,Pot17}. We consider the scattering of a three-body projectile ($a=b+x$, with $b=b_1+b_2$) off a target $A$. We assume that $x$ is structureless and that the $b+A$ interaction is described by some optical potential, while the $x+A$ interaction is given by a microscopic potential. Therefore, the full Hamiltonian of the system is expressed as 
\begin{align}
H &= T_{aA} + V_{xA}(\vec{r}_{x},\xi_A) +  U_{bA}(\vecr_{bA}) + T_{bx
} + V_{bx}(\vec{r}_{bx},\xi_b) ,
\end{align}
where $\xi_b$ and $\xi_A$ denote the degrees of freedom of the core and target systems. 

The differential breakup cross section for the population of an excited state of the $x+A$ system is given by 
\begin{align}
\label{eq:Tprior}
\frac{d^2\sigma}{d \Omega_b \, d E_{b}} = & \normsig \sum_{f} \delta(E_i - E_f) \nonumber \\ & \times | \langle \chi^{(-)}_b(\vec{k}_b) \phi_b(\xi_b) \Phi^f_{xA} | V_\mathrm{post} | \Xi^{(+)}  \rangle |^2 
\end{align}
where $\rho(E_b)$ is the density of states of $b$ particles,  $\phi_b(\xi_b)$  is the core wavefunction, $V_\mathrm{post}= V_{bx}+ U_{bA}-U_{bB}$ with $U_{bB}$ an auxiliary and, in principle, arbitrary potential generating the distorted wave $\chi^{(-)}_{b}$,  $\Phi^f_{xA}$ are the states of the $x+A$ system, and $\Xi^{(+)}$ is the exact solution, with $a+A$ incident channel, of the full model Hamiltonian. The delta function ensures energy conservation, with $E_i=E_f= E_b + \varepsilon^f_{xA} \equiv E$. 

IAV make use of the DWBA approximation, so that the total wavefunction is approximated as  $\Xi^{(+)}\approx \chi^{(+)}_a(\vecr_{a}) \phi_{a}(\vecr_{bx},\xi_b) \phi_A(\xi_A)$, where  $\phi_{a}(\vecr_{bx},\xi_b)$ and  $\phi_A(\xi_A)$ represent the ground state wavefunctions of the projectile and target, respectively, and $\chi^{(+)}_a(\vecr_{a})$ is a distorted wave, solution of a potential $U_a$, typically describing $a+A$ elastic scattering. 

By expressing the energy-conserving delta function in Eq.~(\ref{eq:Tprior}) as the imaginary part of an energy denominator, the sum over final $x+A$ states can be performed using completeness of these states, leading to a closed-form expression for the inclusive breakup cross section. For many applications, it is convenient to separate the elastic and nonelastic breakup (NEB) cross sections, distinguishing the situation in which the target is left in its ground state from the case in which it undergoes some kind of excitation. This separation can be formally performed using the techniques proposed by Kasano and Ichimura \cite{Kas82}. The final formula for the NEB part results:
\begin{align}
\label{eq:iav}
\frac{d^2\sigma^\mathrm{NEB}}{d \Omega_b \, d E_{b}} & 
 = -\normsig    \langle \varphi_x(\vec{k}_b) | W_{xA} | \varphi_x (\vec{k}_b)\rangle
\end{align}
where $v_i$ is the relative velocity in the entrance channel,  $W_{xA}$ is the imaginary part of the $x+A$ optical potential $U_{xA}$ and 
$\varphi_x(\vec{k}_b, \vecr_x)$ the   $x$-channel wavefunction
\begin{align}
    \varphi_x(\vec{k}_b, \vecr_x) = G^{\mathrm{opt}(+)}_{xA}
 \langle \vec{r}_x \phi_b \chi^{(-)}_{b}(\vec{k}_b)  |  V_\mathrm{post} |  \chi^{(+)}_a \phi_a      \rangle \equiv  G^{\mathrm{opt}(+)}_{xA} \rho_x(\vec{k}_b,\vecr_x)
\end{align}
with 
\begin{align}
G^\mathrm{opt(+)}_{xA}(E-E_b) =   \frac{1}{ (E^+ - E_b- T_{xA}-U_{xA})}  
\end{align}
and $\rho_x$ the source term
\begin{align}
\rho_x(\vec{k}_b,\vec r_x)= \langle \vec{r}_x \phi_b \chi^{(-)}_{b} (\vec{k}_b)  |  V_\mathrm{post} |  \chi^{(+)}_a \phi_a      \rangle .
\end{align}

The NEB accounts for breakup processes that involve any kind of excitation of the $x+A$ system, including transfer or the formation of some $A+x$ compound nucleus. In the present work, we are interested in processes in which the $x+A$ system is left in a bound state (i.e., transfer). In that case, the $x$-channel $\varphi_x(\vec{k}_b, \vecr_x)$ function obeys the usual asymptotic decaying form for bound states. 

We now deal with the degrees of freedom of the core, which are present in $\phi_b$, in $V_\mathrm{post}$ (through $V_{bx}$) and also in the total wavefunction $\Xi^{(+)}$. At this point, we make the simplifying assumption that the transition operator $V_\mathrm{post}$ does not alter the state of the $b$ system, in which case the channel wavefunction can be rewritten as 
\begin{align}
\varphi_x(\vec{k}_b, \vecr_x) = G^{\mathrm{opt}(+)}_{xA}
 \langle \vec{r}_x \, \chi^{(-)}_{b} (\vec{k}_b)  |  V_\mathrm{post} | \chi_a f_{ab}(\vec{r}_{bx})  \rangle  
\end{align}
where we have introduced the overlap function between the projectile and the residual nucleus after the particle transfer,
\begin{equation}
    f_{ab}(\vec{r}_{bx})\equiv \int d\xi_b \, \phi^*_b(\xi_b)\phi_a(\vec{r}_{bx},\xi_b).
\end{equation}
In the present work, the projectile is $a\equiv{^9\text{Be}}$ in its ground state, while the ejectile $b\equiv{^8\text{Be}}$ is unbound with respect to the alpha emission and can be found in a series of continuum states characterized by its relative angular momentum $L$ and the relative energy $E_{\alpha\alpha}$.

The transition operator contains the potential $V_{bx}$, which is not well defined in the three-body model of $a$. It is therefore convenient to transform the above expression to its prior-form counterpart, in which this potential does not appear explicitly in the matrix element. This is accomplished by using the following identity (see e.g., Eq.~(9a) of Ref.~\cite{Li84}  or Eq.~(4.18) of Ref.~\cite{IAV85}):  
\begin{equation}
\label{eq:LUT}
\psixpost(\vec{k}_b,\vec r_x)  = \psixprior(\vec{k}_b, \vec r_x)  +  \psixno(\vec{k}_b, \vec r_x)  .
\end{equation}
where
\begin{equation}
\label{eq:psixno}
\psixno(\vec{k}_b,\vec{r}_x) = \langle \vecr_x | \psixno (\vec{k}_b) \rangle =   \langle \vecr_x \chi^{(-)}_b (\vec{k}_b) |  \chi_a^{(+)} \, f_{ab}  \rangle .
\end{equation}
and 
\begin{equation}
\psixprior(\vec{k}_b,\vec r_x) = G^{\mathrm{opt}(+)}_{xA}
 \langle \vec{r}_x \, \chi^{(-)}_{b}(\vec{k}_b)  |  V_\mathrm{prior} | \chi^{(+)}_a f_{ab}(\vec{r}_{bx})  \rangle  
\end{equation}
with $V_\mathrm{prior} \equiv U_{xA} + U_{bA}-U_{a}$. 
The NEB double differential cross section for the detection of $b$ is then calculated by inserting \ref{eq:LUT} into Eq.~(\ref{eq:iav}).


\begin{figure}[ht]
\centering
\includegraphics[width=0.9\linewidth]{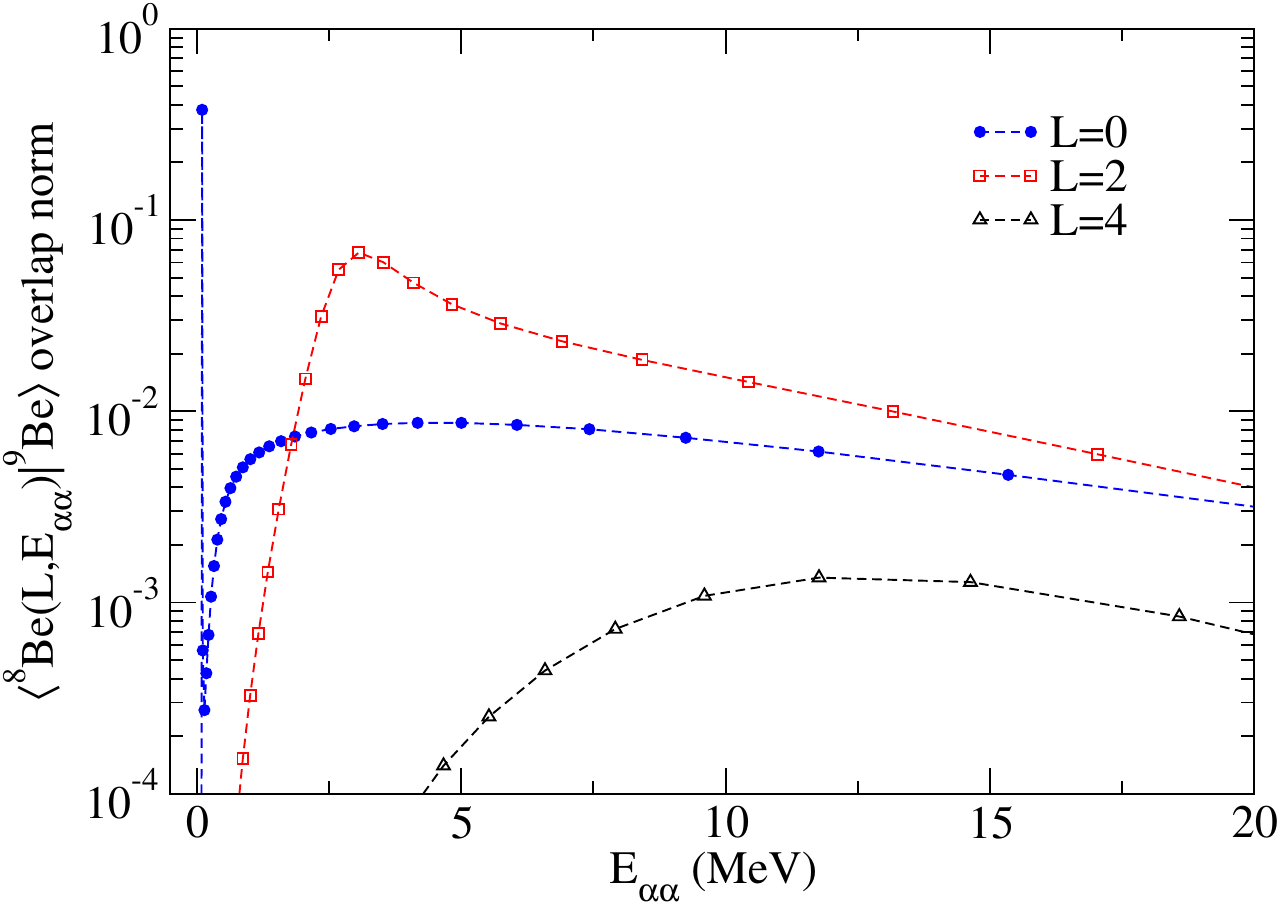}
\caption{\label{fig:strength} Norm of the overlaps between the $^9$Be g.s. and the $^8$Be continuum pseudostates for different relative angular momenta $L$, as a function of the $\alpha$-$\alpha$ relative energy up to 20 MeV. Dashed lines are included as a guide.}
\end{figure}

\subsection{Structure input}
The present calculations require a description of the $^{9}$Be ground state, the $^8$Be continuum states and their corresponding overlaps. Due to the Borromean structure of $^9$Be, these can be obtained within a three-body ($\alpha+\alpha+n$) model with effective pairwise interactions, provided the same $\alpha+\alpha$ potential is used to generate the two-body $^8$Be states. In this work, we adopt the $^9$Be model of Refs.~\cite{cas14,cas15}, which is built within the hyperspherical harmonics expansion method.
The $\alpha$-$n$ interaction is modeled with the Woods-Saxon potential in Ref.~\cite{thom00}, including central and spin-orbit terms, while the Ali-Bodmer potential is employed for the $\alpha$-$\alpha$ interaction~\cite{AliBodmer}. Both nuclear potentials include repulsive-core components to account for the Pauli principle. The electrostatic repulsion between the $\alpha$'s is described with a hard-spheres Coulomb interaction, and an additional three-body force is included to correctly reproduce the separation energy of $^9$Be ground state. 

Since $^8$Be is unbound, overlaps will be a function of the continuum energy $E_{\alpha\alpha}$. For simplicity, we compute the $^8$Be states using a pseudostate method, i.e., expanding the state in a discrete basis of square-integrable functions. For this purpose, we employ the analytical transformed harmonic oscillator (THO) basis~\cite{Moro09,Lay10}. We obtain $^8$Be pseudostates for relative angular momenta $L=0,2$ and $4$, in a basis of 40 THO functions with parameters $b=1.6$ fm and $\gamma=1.4$ fm$^{1/2}$. Then we compute the overlaps between each pseudostate and the $^9$Be ground state by integrating over the relative $\alpha$-$\alpha$ coordinates, for specific orbital configurations of the neutron. The norm of the overlaps, for the $^9$Be neutron in the $p_{3/2}$ orbital, is shown in Fig.~\ref{fig:strength}. The low-lying peak corresponds to the $0^+$ ground-state resonance of $^8$Be at $\sim$0.10 MeV above threshold, which exhausts $\sim$38\% of the total $^9$Be norm. The remaining $0^+$ states add up to 53\% of the norm. These weights can be regarded as spectroscopic factors within the three-body model. In turn, the weight of the $2^+$ and $4^+$ states is 44\% and 1\%, respectively. The former is concentrated around 3 MeV, i.e., the nominal energy of the  $2_1^+$ resonance in $^9$Be. The latter is rather small, so 4$^+$ states will not be considered for the reaction calculations. Note that other neutron orbital configurations in the wave function are negligible compared to those considered here.

\begin{figure}[ht]
\centering
\includegraphics[width=0.9\linewidth]{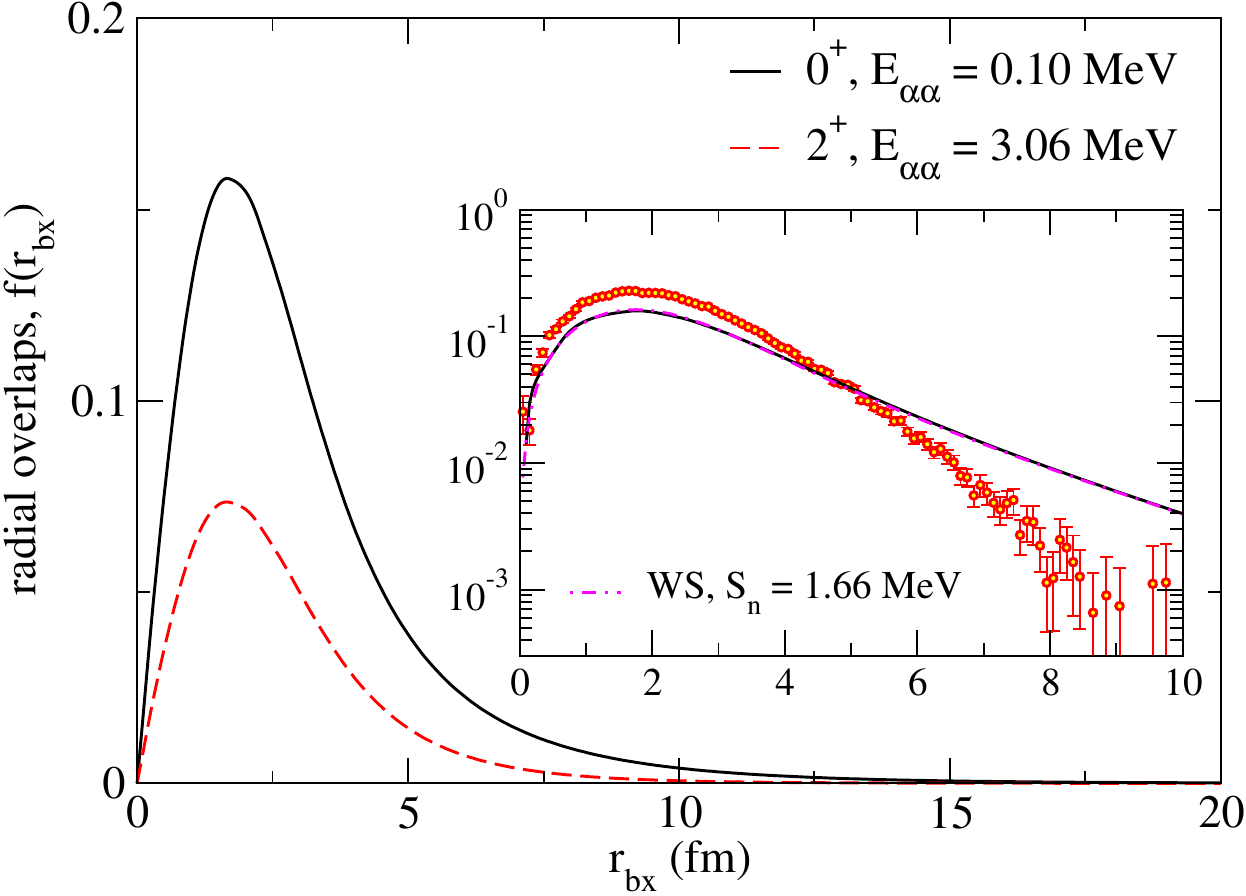}
\caption{\label{fig:radial} Radial overlaps for specific $^8$Be pseudostates: $0^+$ at 0.10 MeV (black solid), $2^+$ at 3.06 MeV (red dashed). Inset: $0^+$ overlap, in logarithmic scale, compared with VMC data for the $0^+$ (see the text).}
\end{figure}

Some characteristic radial overlaps, for $^8$Be pseudostates around the $0^+$ and $2^+$ nominal resonance energies, are shown in Fig.~\ref{fig:radial}. The latter presents a smaller norm, since the $2^+$ strength is more evenly distributed according to Fig.~\ref{fig:strength}. For comparison, the inset also shows the microscopic overlap from variational Monte Carlo calculations (VMC) for the $0^+$ state, which corresponds to a spectroscopic factor (SF) of 0.595~\cite{VMC_9-10,vmc_tables}, and with a phenomenological overlap obtained with a Woods-Saxon potential ($R_0=1.25 \times 9^{1/3}$~fm, $a=0.65$ fm), whose depth is fixed to get a separation energy of $S_n=1.665$~MeV. When adjusted to have the same norm, it is shown that the present $0^+$ overlap matches the correct asymptotic behavior, in contrast to the VMC one, which typically has problems describing the wave function tails. It should be noted that
our overlap corresponds to a single $0^+$ pseudostate (with norm 0.38) and not the whole $0^+$ continuum, thus its magnitude is smaller than that of the VMC one.
Note also that our total norm is smaller than the VMC SF, and this may have an effect on the cross sections, as will be discussed later.

\subsection{Choice of interactions}
The IAV formula (\ref{eq:iav}) requires optical potentials for the entrance ($^{9}$Be+$^{197}$Au) and exit ($^{8}$Be+$^{198}$Au) channels, as well as for the core-core system appearing in the transition operator ($^{8}$Be+$^{198}$Au). All these interactions are taken from the double-folding global Sao Paulo potential (SPP) \cite{Cha02}, with the required densities parametrized in terms of two-parameter Fermi distributions with parameters obtained from a systematic fit of Dirac-Hartree-Bogoliubov calculations and  experimental electron scattering data. For the imaginary part, we used the same radial dependence as the real part, scaled by a standard renormalization factor $N_i=0.78$ \cite{Alv03}. In the $^{8}\text{Be}+{^{197,198}}\text{Au}$ cases, the optical potential may not be well defined due to the unbound nature of $^{8}$Be. As we show below, most of the calculated cross section is due to the $^{8}$Be(g.s.), which corresponds to a very narrow resonance in which the two alphas remain highly correlated for a long time before decaying. This may serve as a {\it posteriori} justification for the use of an optical potential for this system. The calculation requires also the neutron-target interaction at negative neutron energies. In the spirit of the IAV formalism this potential should be dispersive so that its real part should account for the centroids of the relevant single-particle configurations whereas the imaginary part should describe the fragmentation of the single-particle strength due to beyond-mean-field effects, such as the coupling with core-excited states of $^{197}$Au  \cite{Mah85,Mah91}. According to a Hartree-Fock calculation with the SkX interaction \cite{SkX}, the bound single-particle configurations above the Fermi level are: $1i_{11/2}$ ($\epsilon=-0,97$~MeV), $2g_{9/2}$ (-2.46~MeV), $2g_{7/2}$ (-0.255~MeV), $3d_{5/2}$ (-1.13~MeV), $3d_{3/2}$  (-0.42~MeV) and $4s_{1/2}$ (-0.907~MeV). 
We found that the Koning-Delaroche global potential~\cite{Kon03} predicts the same bound-state configurations, with differences in the energies of less than 1~MeV. Therefore, we have adopted this potential for the reaction calculations. It is worth noting that, to reduce the computational demands, we neglected the intrinsic spin of the transferred neutron, so the spin-orbit partners collapse into a single configuration located at their centroid.


\section{Results}\label{sec:results}

\begin{figure}[ht]
\centering
\includegraphics[width=0.9\linewidth]{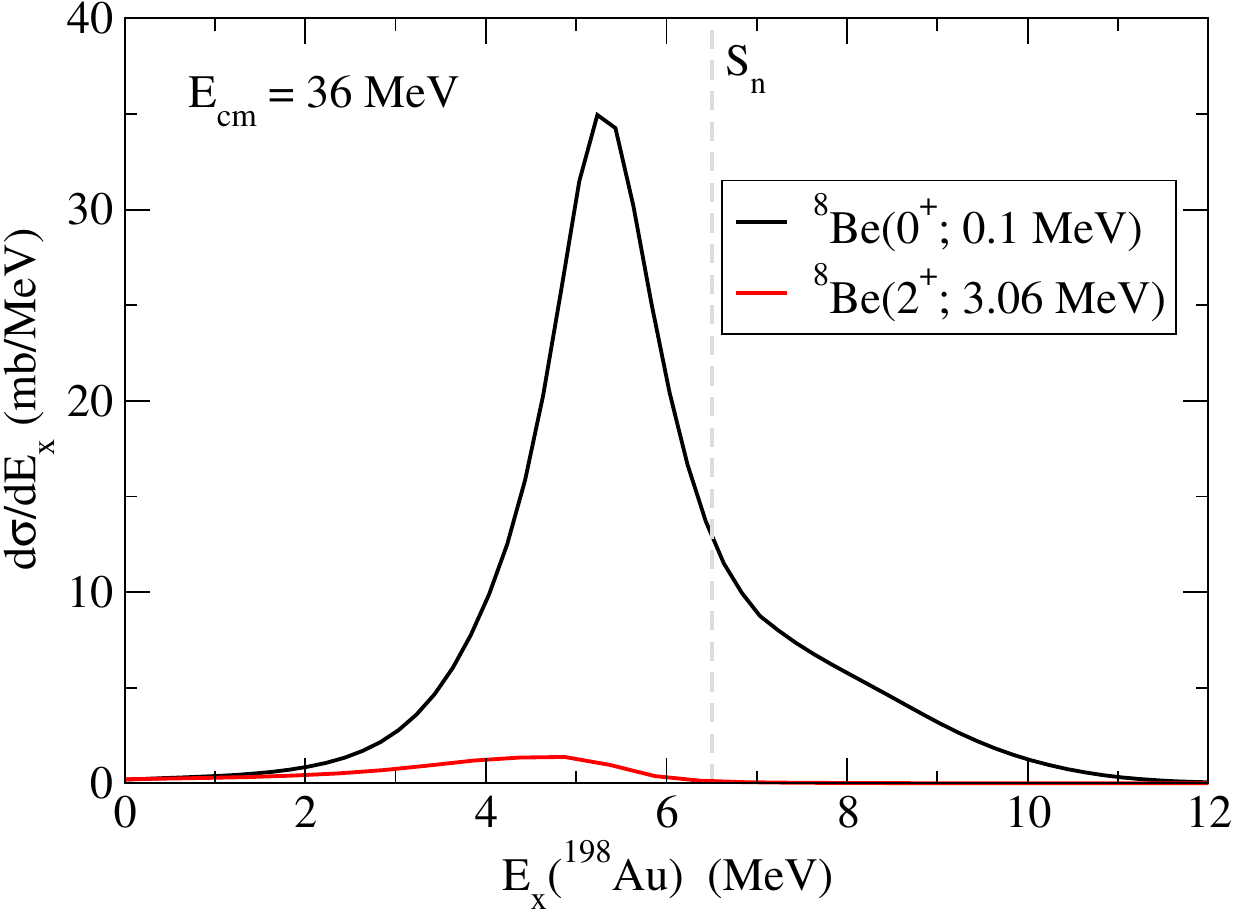}
\caption{\label{fig:dsdex_e36} Differential energy cross section at $E_\mathrm{cm}=36$ MeV, as a function of the $^{198}$Au excitation energy, for two final states of the $^{8}$Be$^*$ system: the $0^+$ ground-state and a state around the nominal energy of the first $2^+$ resonance. The vertical dashed line marks the neutron separation energy in $^{198}$Au.}
\end{figure}

\begin{figure}[ht]
\centering
\includegraphics[width=0.9\linewidth]{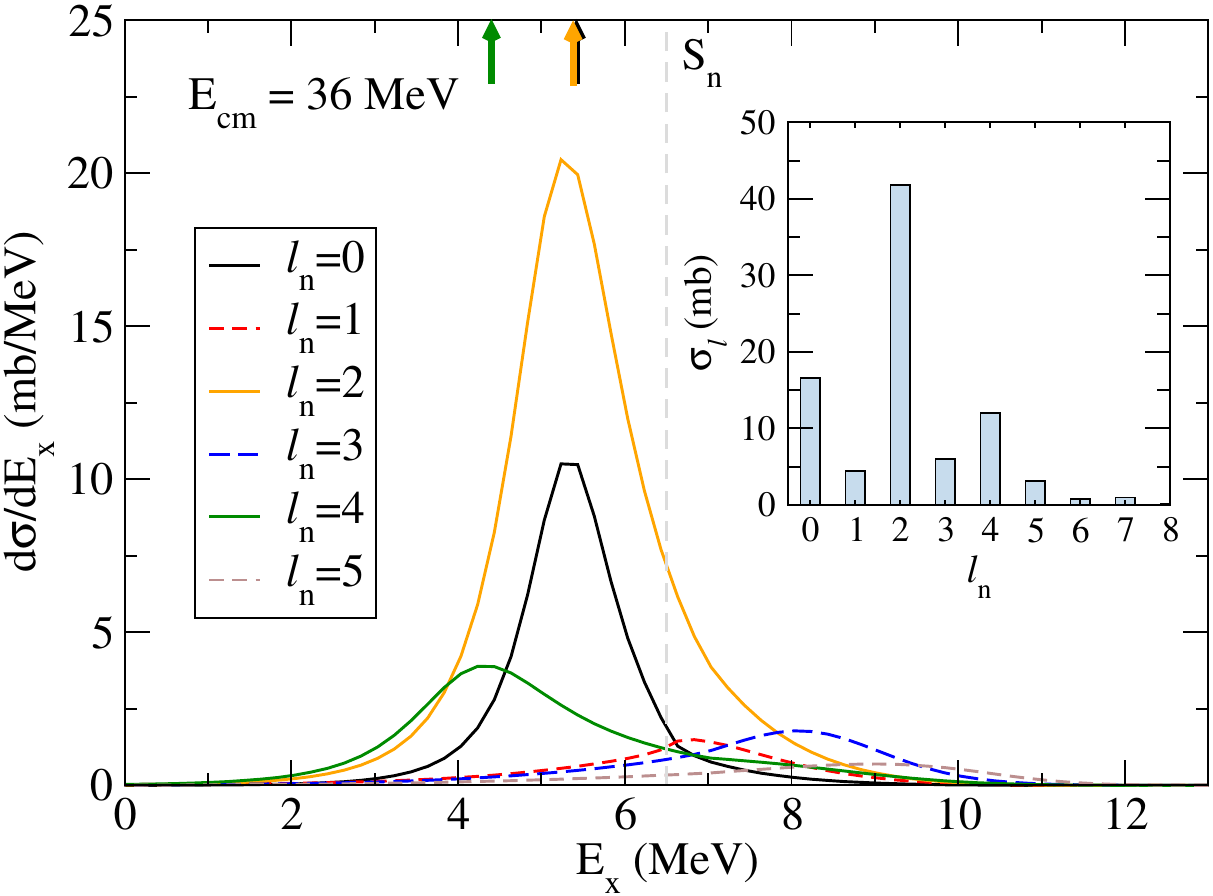}
\caption{\label{fig:dsdex_e36_lx} Decomposition of energy differential cross section at $E_\mathrm{cm}=36$ MeV, for $^8$Be in its 0$^+$ ground state, in terms of the relative orbital angular momentum between the transferred neutron and the target. The inset shows the integrated contribution of each $n+{^{197}}$Au  orbital angular momentum. The vertical arrows indicate the position of the single-particle states of the $n+{^{197}}$Au potential.}
\end{figure}

The measured data are doubly inclusive with respect to the final state of the residual nucleus ($^{198}$Au) as well as with respect to the state of the outgoing $^8$Be$^*$ (i.e., $2 \alpha$) system. Expression (\ref{eq:iav}) provides the energy distribution of the residual nucleus for a given internal state of the $^{8}$Be core. Therefore, for a meaningful comparison with the data, one needs to evaluate this expression for a range of $^{8}$Be states including all relevant relative angular momenta and energies between the two $\alpha$'s. Due to the discretization of the $^{8}$Be states, this translates into a sum of the contribution for the different pseudostates. For illustrative purposes we show in Fig.~\ref{fig:dsdex_e36} the differential cross section, at an incident energy of $E_\mathrm{cm}=36$ MeV, as a function of the excitation energy of the $^{198}$Au system for two selected pseudo-states of $^{8}$Be: a $0^+$ characterizing the ground state of the system and a $2^+$ pseudostate around the nominal energy of the lowest $2^+$ state. It is seen that the cross section is largely dominated by the former, that is, the two $\alpha$'s are preferably emitted in its ground-state resonance. Note that for a meaningful comparison with the data, only the bound state region of the $^{198}$Au  system must be considered, so the differential cross section must be integrated from $E_x=0$ to $E_x=S_n$ (indicated by a vertical dashed line in the plot). The partial-wave decomposition in terms of the relative angular momentum $l_n$ between the transferred neutron and the $^{197}$Au target is shown in Fig.~\ref{fig:dsdex_e36_lx} for the case in which $^8$Be is produced in its ground state resonance. It is seen that $l_n=0,2$ and $4$ dominate the cross section, with peaks around the single-particle bound states of the adopted $n+{^{197}}$Au potential (located at $E_x=5.42$, 5.38 and 4.39~MeV, indicated by the arrows), respectively. Note that within the IAV scheme, the single-particle strength is spread in energy and some part extends into the unbound region (i.e.\ $E_x > S_n$). 
Notice also that, with the adopted potential, most of the odd-$l_{n}$ strength lays also in the continuum and will not contribute to the cross section of interest. 

\begin{figure}[ht]
\centering
\includegraphics[width=0.9\linewidth]{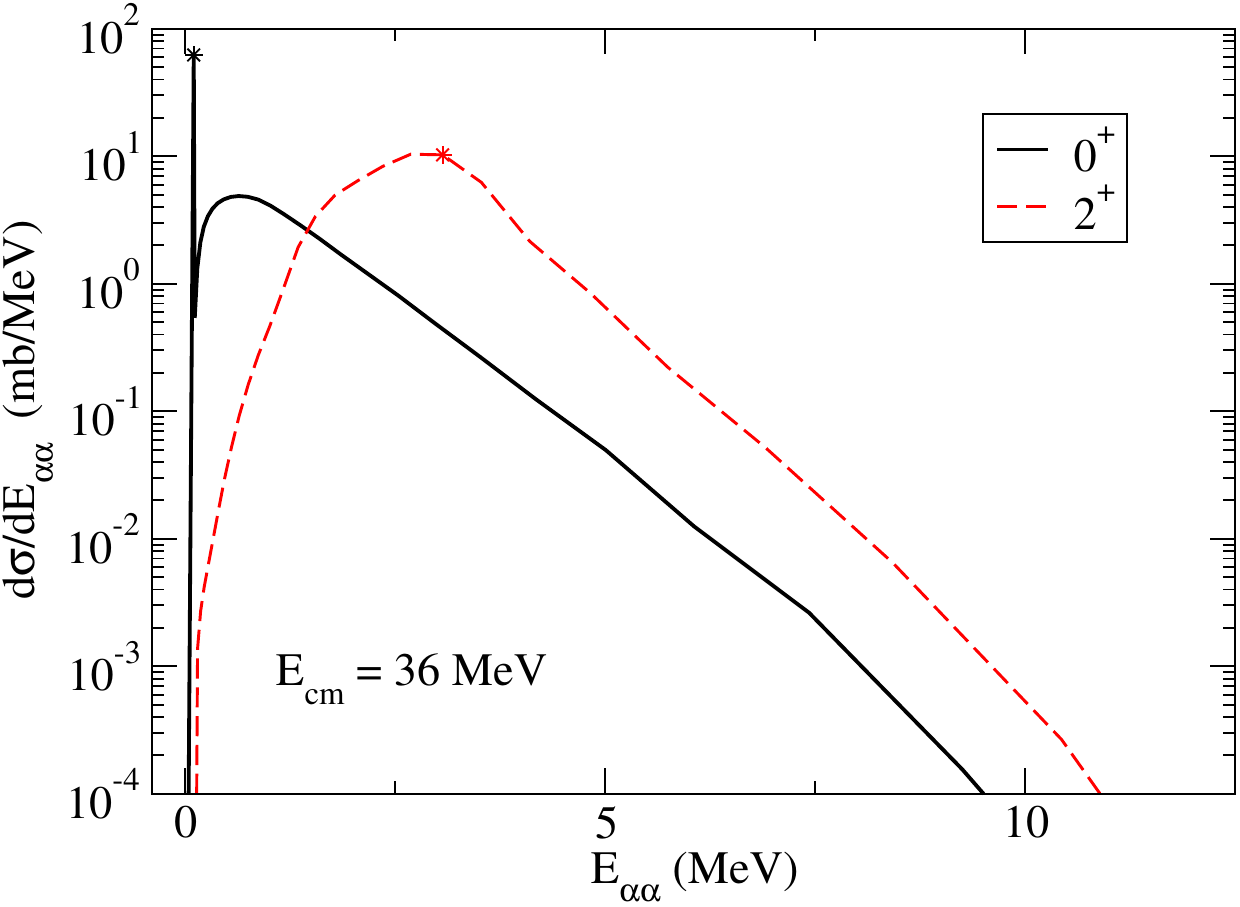}
\caption{\label{fig:dsde8Be} Differential cross section, at $E_\mathrm{cm}=36$ MeV, as a function of the $\alpha$-$\alpha$ relative energy in the final $^8$Be$^*$ system for 0$^+$ states (solid line) and $2^+$ states (dashed line). The states selected for Fig.~\ref{fig:dsdex_e36} are highlighted with a symbol.}
\end{figure}

As already stated, the procedure should be repeated for all pseudostates considered in the structure calculation of $^8$Be. In Fig.~\ref{fig:dsde8Be} we show the differential cross section as a function of $E_{\alpha-\alpha}$. Note that, in a pseudostate description, we obtain a series of discrete cross section values for each final $^8$ Be state, and then these results are translated into a continuous differential cross section following the procedure used, for instance, in \cite{Die17}. 
This distribution reflects, in part, the weights of the different $^8$Be configurations in the $^9$Be projectile wave function displayed in Fig.~\ref{fig:strength}, but also the reaction dynamics that seem to favor the population of the $0^+$ ground-state resonance over other states.

\begin{figure}[ht]
\centering
\includegraphics[width=0.9\linewidth]{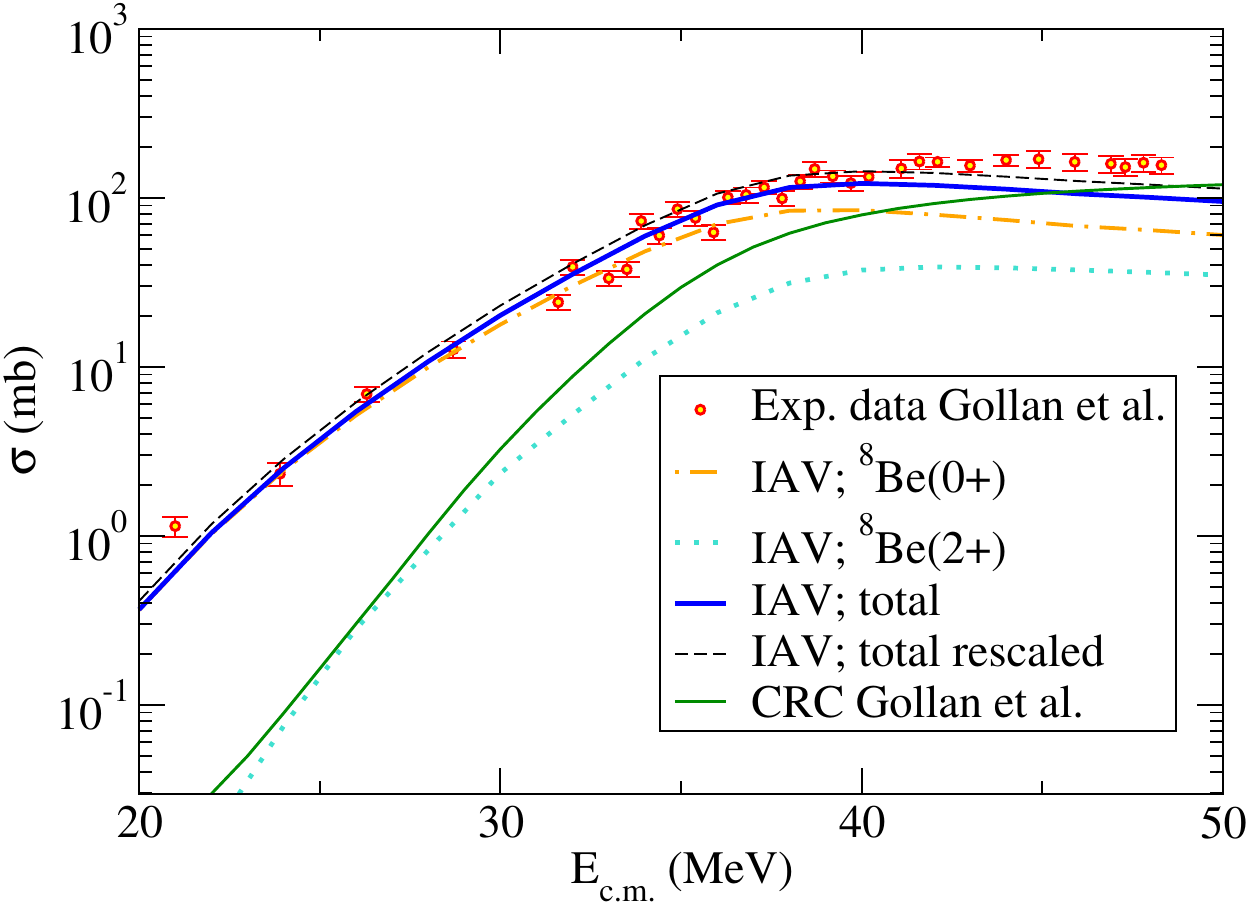}
\caption{\label{fig:final2} Angle-integrated total transfer cross section as a function of $E_\mathrm{cm}$. The contribution from the $0^+$ (dot-dashed line) and $2^+$ (dotted) states of $^8$Be, together with their sum (blue solid), are compared to the experimental data and the CRC calculation (green solid) of Ref.~\cite{gollan21}. The IAV result rescaled by the VMC spectroscopic factors (dashed) is also shown (see text).}
\end{figure}

We then perform the same calculations at different incident energies from $E_\mathrm{cm}=20$ to 50 MeV, in order to compare with the experimental data of Ref.~\cite{gollan21}. The results are presented in Fig.~\ref{fig:final2}. It is seen that the $0^+$ states of $^8$Be dominate over a wide range of incident energies, and that the $2^+$ contribution is significant only above $E_\mathrm{cm}=40$ MeV. 
Adding the two contributions, the cross section matches reasonably well both the trend and magnitude of the experimental data, except for a noticeable underestimation at the highest energies considered. Here, it is worth noting that the structure of $^9$Be is computed within an effective three-body model that can only treat the Pauli principle approximately. As a consequence, the weights of our different $^8$Be configurations add up to unity, while the SF coming from microscopic calculations such as VMC, considering actual antisymmetrization and occupation numbers, are larger. Thus, we may tentatively rescale our calculations by the VMC spectroscopic factors of 0.595 and 0.603 for the $0^+$ and $2^+$ states, respectively. The results are also shown in Fig.~\ref{fig:final2}, where an improvement is obtained in the region where the $2^+$ contribution is important, due to a larger difference between the three-body weight and the corresponding VMC SF. Nevertheless, the present IAV calculations with structure overlaps within an effective three-body model can explain the main features of the experimental cross section. 
In the figure we show also the coupled-reaction-channel (CRC) calculations presented in the original experimental work~\cite{gollan21}. In this calculation, only states of $^{198}$Au up to $\sim 700$~keV excitation energy included, due to computational limitations. The result was not able to describe the inclusive transfer data. A possible explanation for the difference with our IAV result is the missing strength up to the separation threshold of $^{198}$Au ($S_n=6.5$~MeV), which we include effectively with the choice of the $n$-${^{197}}$Au optical potential. 
We have also compared our results with those reported by Kaushik {\it et al.}~\cite{kaushik21}, where the same reaction was measured and compared with CRC calculations. In these calculations, the authors include states of the residual nucleus up to $E_x=1.56$~MeV (assuming unit spectroscopic factors for all states) and obtain a good agreement with the ($^9$Be,$^8$Be) data, with a dominance of the  $^{8}$Be($2^+$) state in the full energy range. Our test calculations suggest, however, that both effects (the seemingly good agreement with the data and the $^{8}$Be($2^+$) dominance) might be a byproduct of the unjustified assumption of unit spectroscopic factors for the considered states, and the omission of higher exited states of $^{198}$Au.

\begin{figure}[ht]
\centering
\includegraphics[width=0.9\linewidth]{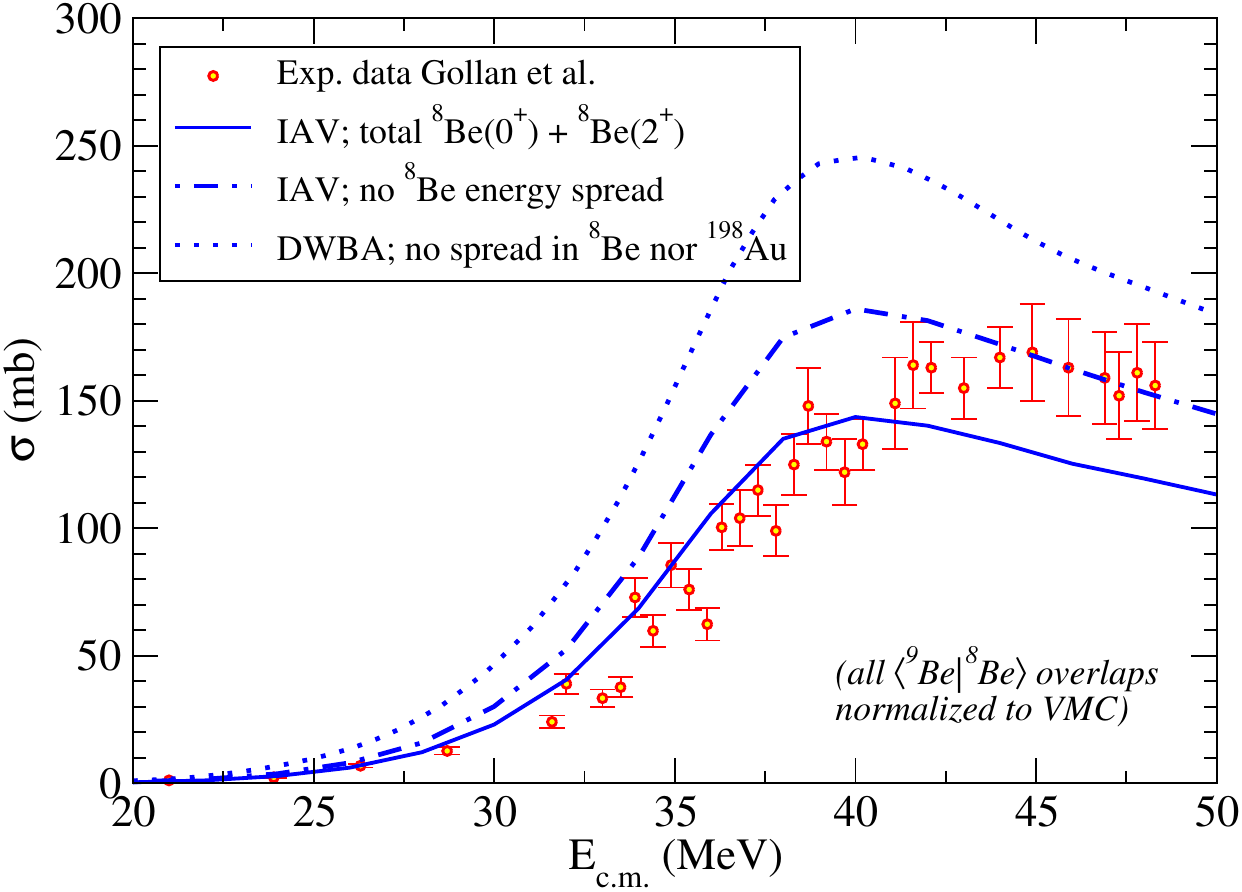}
\caption{\label{fig:no-frag} Same as Fig.~\ref{fig:final2}, but comparing the full IAV calculation (solid line) with IAV and standard DWBA calculations in which one ignores the energy spread of the $^{8}$Be states and $^{198}$Au states in the overlaps  $\langle \mathrm{^8Be}| \mathrm{^{9}Be}\rangle$ and  $\langle \mathrm{^{198}Au}| \mathrm{^{197}Au}\rangle$, respectively. All overlaps are normalized to the prediction of the VMC calculation of \cite{VMC_9-10}. }
\end{figure}

As noted above, the two main novelties of the present approach with respect to more standard DWBA calculations (or their extended CCBA and CRC versions) is the consideration of the energy distribution of the $^{8}$Be and $^{198}$Au states which enter the reaction calculations through the overlap functions $\langle \mathrm{^8Be}| \mathrm{^{9}Be}\rangle$ and $\langle \mathrm{^{198}Au}| \mathrm{^{197}Au}\rangle$, respectively.  It is therefore instructive to compare these results with those obtained from standard DWBA calculations, neglecting the effect of the energy distribution of these overlaps, that is, in which the full strength for each overlap is concentrated in a single state with a definite energy. The results of these test calculations are summarized in Fig.~\ref{fig:no-frag}, which is presented in linear scale to better highlight the differences between the different calculations. The blue solid line is the original IAV calculation, in which the energy spreading of the structure overlaps is considered for both the $^{8}$Be and $^{198}$Au states. In the dot-dashed line, the $\langle \mathrm{^8Be}| \mathrm{^{9}Be}\rangle$  overlaps of the $0^+$ and $2^+$ states of $^{8}$Be are represented by the pseudostates closest to the nominal energies of the associated resonances, i.e., 0.1~MeV and 3.06~MeV, and renormalized to give the SF predicted by the VMC calculation. An apparent increase is observed with respect to the original IAV calculation, overestimating the data for energies below $E_\mathrm{cm}=42$~MeV. Finally, the dotted line is the calculation in which the energy distribution of both overlaps is neglected. In $^{198}$Au, we consider a pure single-particle state for each $nlj$ configuration located at the nominal energy predicted by the real part of the KD potential. Restricting to the bound-state configurations, only the $l_n$=0,2,4 and 6 waves are considered. As in the IAV calculations, the spin-orbit term is not considered, so these configurations are degenerate in $j_n$ (where $\vec{j}_n=\vec{l}_n +  \vec{s}$). This calculation produces a further increase in the calculated transfer cross section, overestimating the data in the full energy range. Therefore, from these calculations, we conclude that consideration of the energy spread of the structure overlaps is an important ingredient for a quantitative description of the experimental data. 
A question remains about the apparent underestimation at the highest energies considered. A possible reason for this difference may be attributed to configurations of the $^9$Be nucleus not explicitly considered here, and this calls for further theoretical investigations. 


\section{\label{sec:sum} Summary and conclusions}
In summary, we have presented a new methodology to evaluate inclusive transfer cross sections induced by a Borromean nucleus like $^{9}$Be. We have considered the stripping reaction    $^{197}$Au($^{9}$Be,$^{8}$Be)X which has been recently measured at energies below and above the Coulomb barrier. The proposed strategy incorporates two key ingredients with respect to more standard approaches. First, the dense spectrum of the final $^{198}$ Au states is described approximately using a complex neutron + target potential, whose imaginary part accounts for the energy distribution of the single-particle strengths. Second, the required $\langle \mathrm{^8Be}| \mathrm{^{9}Be}\rangle$ overlaps, including unbound states of $^8$Be, are computed with a three-body model for the $^9$Be projectile. This, again, permits the incorporation of the energy spread of the $^8$Be states. 

Aside from providing a more consistent treatment of the structure and reaction dynamics, the calculations based on this new methodology describe better the available experimental data for the studied reaction, as compared to more standard approaches. Moreover, the present method would allow the evaluation of more detailed cross sections, such as differential cross sections with respect to the scattering angles of the $^{8}$Be system or for the relative energy between the two outgoing alphas \cite{Cook16}.



The present work may pave the way towards a better understanding of the impact of neutron-induced breakup on the fusion cross sections and, in particular, on the above-barrier complete fusion suppression reported for this and other $^{9}$Be-induced reactions.

\section*{Acknowledgements}
We are grateful to Hugo Soler, Jes\'us Lubian and Vivek Parkar for enlightening discussions about the CRC calculations employed in previous works. This work has been partially funded by MCIN/AEI/10.13039/501100011033 under I+D+i project No.\ PID2020-114687GB-I00. J.L. is supported by the National Natural Science Foundation of China (Grants No.12105204), by the Fundamental Research Funds for the Central Universities.

\bibliography{newbibfile}

\end{document}